\documentclass[aps,prc,preprint,amsmath,showpacs,superscriptaddress,nofootinbib]{revtex4-1}
\newcommand{\beqy}{\begin{eqnarray}}
\newcommand{\eeqy}{\end{eqnarray}}
\newcommand{\bmlet}{\begin{subequations}}
\newcommand{\emlet}{\end{subequations}}
\usepackage{graphicx}
\usepackage{dcolumn}
\usepackage{bm}
\usepackage{hyperref}
\usepackage[colorinlistoftodos]{todonotes}

\def\apj{ApJ}  
\def\apjl{ApJ} 
\def\prl{Phys.~Rev.~Lett.} 
\def\pr{Phys.~Rev.} 
\def\prc{Phys.~Rev.~C} 
\def\prx{Phys.~Rev.~X} 
\def\aap{A\&A}   
\def\physrep{Phys.~Rep.}   
\def\nphysa{Nucl.~Phys.~A}   
\def\mnras{MNRAS}             
\def\epja{Eur.~Phys.~J.~A}
\def\cpc{Chin.~Phys.~C}
\def\jpgnpp{J.~Phys.~G:~Nucl.~Part.~Phys.} 
\def\plb{Phys.~Lett.~B} 

\begin{document}

\title{Role of the symmetry energy and the neutron-matter stiffness on the tidal deformability of a neutron star with unified equations of state}

\author{L. Perot}
\affiliation{Institute of Astronomy and Astrophysics, Universit\'e Libre de Bruxelles, CP 226, Boulevard du Triomphe, B-1050 Brussels, Belgium}
\author{N. Chamel}
\affiliation{Institute of Astronomy and Astrophysics, Universit\'e Libre de Bruxelles, CP 226, Boulevard du Triomphe, B-1050 Brussels, Belgium}
\author{A. Sourie}
\affiliation{Institute of Astronomy and Astrophysics, Universit\'e Libre de Bruxelles, CP 226, Boulevard du Triomphe, B-1050 Brussels, Belgium}
\affiliation{LUTH, Observatoire de Paris, PSL Research University, CNRS, Universit\'e Paris Diderot,  Sorbonne Paris Cit\'e, 5 place Jules Janssen, 92195 Meudon, France.}

\begin{abstract}
The role of the symmetry energy and the neutron-matter stiffness on the tidal deformability of a cold nonaccreted neutron star is studied using a set of unified equations of state. 
Based on the nuclear energy-density functional theory, these equations of state provide 
a thermodynamically consistent treatment of all regions of the star and were calculated 
using functionals that were precision fitted to experimental and theoretical nuclear data. 
Predictions are compared to constraints inferred from the recent detection of the gravitational-wave signal GW170817 from a binary neutron-star merger and from observations of the electromagnetic counterparts. 
\end{abstract}

\maketitle

\section{Introduction}
\label{sec:intro}

The detection of the gravitational-wave signal GW170817 from the merger of two neutron stars (NSs)~\cite{ligo2017inspiral} and the subsequent observations of electromagnetic counterparts~\cite{ligo17gwgrb, goldstein17,  ligo17multi, coulter17, troja17, haggard17, hallinan17} offer new opportunities to probe the properties of matter under conditions so extreme that they 
cannot be experimentally reproduced (see, e.g., Ref.~\cite{blaschke2018} for a recent review). Apart from estimates of the masses of the two inspiralling NSs, the analysis of this signal has also provided valuable information on their tidal deformations during the last orbits~\cite{de2018,ligo2018, ligo2019prx}. The relatively small dimensionless tidal deformability (or polarizability) parameter
\begin{equation}
    \label{eq:Lambda}
    \Lambda=\dfrac{2}{3} \,   k_2 \, \left(\dfrac{c^2 R}{GM}\right)^5
\end{equation} 
(with $R$ the circumferential radius of the star, $M$ its mass, and $k_2$ the second gravito-electric Love number, $c$ the speed of light, $G$ the gravitational constant)  inferred from GW170817 has already ruled out the stiffest equations of state (EoSs) of high-density matter, see, e.g., Refs.~\cite{ligo2017inspiral,de2018,ligo2018,ligo2019prx}. Subsequent studies aimed at further examining possible correlations between the tidal deformability of a NS and properties of finite nuclei or infinite nuclear matter such as the symmetry energy, see, e.g., Refs.~\cite{malik2018,fattoyev2018,zhangli2019,zhangli2019b,tong2019,carson2019, krastev2019,tsang2019,tsang2019b,malik2019,raithel2019b,wei2019}. Most studies carried out so far have focused on the NS core, employing different models for the crust (such as the outdated EoSs of Refs.~\cite{bps1971} and \cite{nv1973} for the outer and inner crusts respectively) or merely using polytropic EoSs. 
However, a proper treatment of the crust and a consistent determination of the crust-core boundary is important for reliable calculations of NS radii (see, e.g., Refs.~\cite{lattimer2007,piekarewicz2014,fortin2016}) and tidal Love number $k_2$ ~\cite{piekarewicz2019,kalaitzis2019,gamba2019}, especially for the range of NS masses inferred from GW170817. 

In this paper, the role of dense-matter properties on the tidal deformability of a NS is examined using a series of seven unified EoSs, BSk19, BSk20, BSk21, BSk22, BSk24, BSk25, and BSk26~\cite{potekhin2013,pearson2018}, calculated in the framework of the nuclear energy-density functional theory (see, e.g., Ref.~\cite{duguet2014} for a review). These EoSs, whose main features are recapitulated in Section~\ref{sec:uni_eos}, provide a thermodynamically consistent description of all regions of a NS, from the surface to the central core of the star. The underlying functionals were precision fitted to essentially all experimental atomic mass data and were simultaneously adjusted to theoretical nuclear data.
The series BSk19, BSk20 and BSk21~\cite{gcp2010} were fitted to realistic neutron-matter (NeuM) EoSs with different degrees of stiffness, while the series BSk22, BSk24, and BSk25~\cite{gcp2013} mainly differ in their predictions for the symmetry energy (BSk26 being fitted to the same symmetry-energy coefficient at saturation as BSk24 but to a different NeuM EoS). The corresponding EoSs are thus used to assess the role of the symmetry energy and of the NeuM stiffness on the tidal deformability of a NS in Section~\ref{sec:tidal}. Theoretical predictions are compared to observations of GW170817 in Section~\ref{sec:GW170817}.

\section{Unified equations of state for neutron stars}
\label{sec:uni_eos}

The unified NS EoSs were calculated under the cold-catalyzed matter hypothesis, i.e., electrically charge-neutral matter in its absolute ground state~\cite{hw58,htww65}. These EoSs are based on the Brussels-Montreal functionals, whose main properties are presented in Section~\ref{sec:EDF}. The methods employed to calculate the EoS in the different regions of a NS are briefly reviewed in Section~\ref{sec:model}. 

\subsection{Brussels-Montreal energy-density functionals}
\label{sec:EDF}

The functionals considered here are based on generalized Skyrme effective interactions with terms that depend on both the relative momentum $\pmb{p}_{ij} = - {\rm i}\hbar(\pmb{\nabla}_i-\pmb{\nabla}_j)/2$ of nucleons $i$ and $j$, and the average nucleon number density $n(\pmb{r})$ at position $\pmb{r} = (\pmb{r}_i + \pmb{r}_j)/2$~\cite{cgp2009}. Nuclear pairing is treated using a different effective interaction constructed from realistic $^1S_0$ pairing gaps in NeuM and symmetric nuclear matter (SNM)~\cite{cgp2008,gcp2009,gcp2009b,chamel2010b}. 
The parameters of the series BSk22-BSk26~\cite{gcp2013} were determined primarily by fitting to 
the 2353 measured masses of atomic nuclei having proton number $Z\geq8$ and neutron number $N\geq8$ from the 2012 Atomic Mass Evaluation~\cite{AME2012}. These functionals provide equally good fits to the 2408 measured masses of nuclei with $N,Z \geq 8 $ from the 2016 AME~\cite{AME2016}. Nuclear masses were calculated using the self-consistent Hartree-Fock-Bogoliubov (HFB) method allowing for axial deformations~\cite{samyn2002}. Phenomenological corrections were added to the HFB energy to account for dynamical correlations and Wigner effects (see, e.g., Refs.~\cite{gcp2010,cgp2008} for a discussion). 
Moreover, a correction for the finite size of the proton was made to both the 
charge radius and the energy~\cite{cgp2008}. Finally, Coulomb exchange for protons was 
dropped, thus simulating neglected effects such as Coulomb correlations, charge-symmetry breaking, and vacuum polarization~\cite{gp2008}. 

\begin{figure}[ht]
\begin{center}
\includegraphics[width=0.75\textwidth]{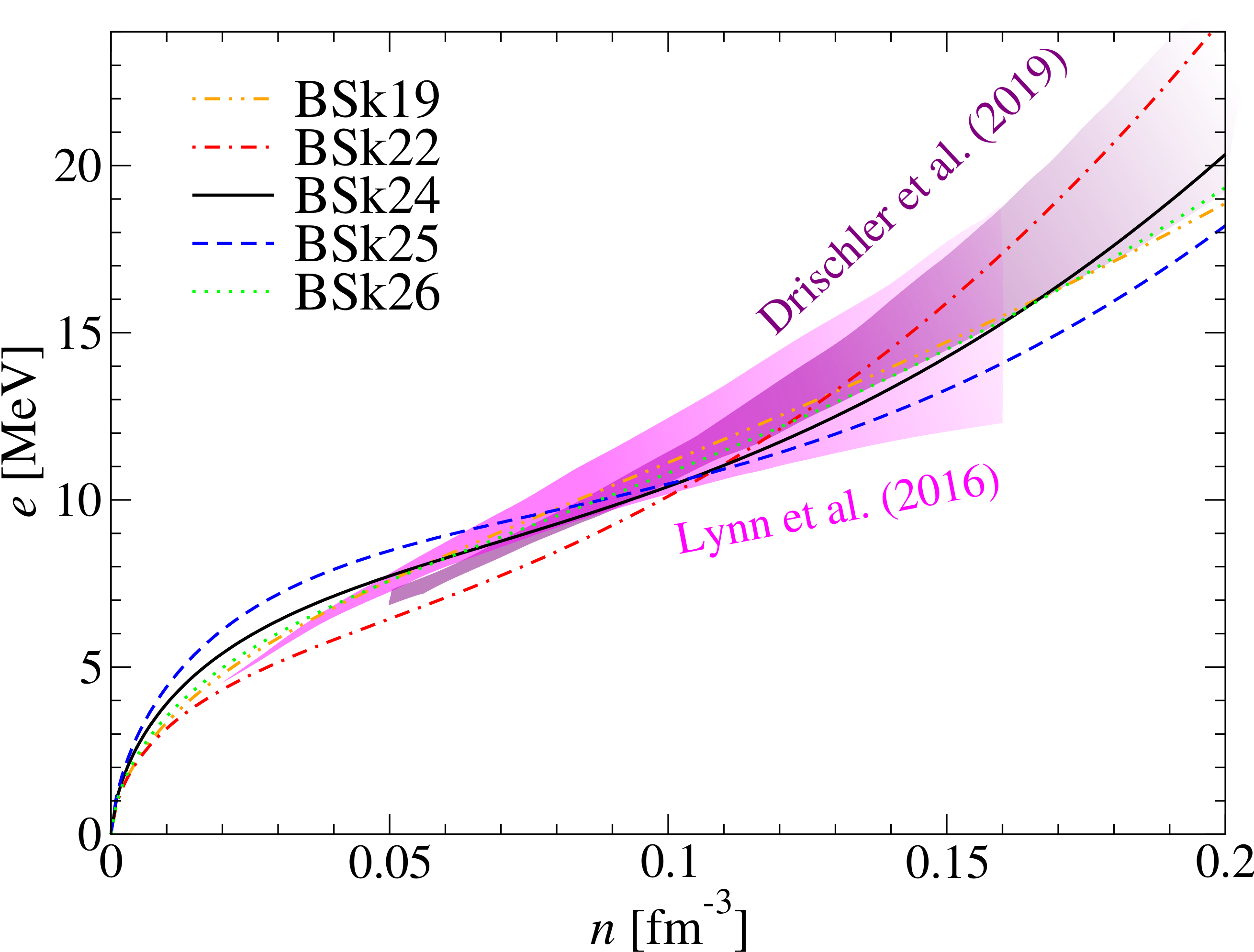}
\caption{(Color online) Energy per particle in NeuM with respect to the number density $n$ as calculated with the functionals BSk19, BSk20, BSk21, BSk22, BSk24, BSk25 and BSk26. The shaded areas represent recent constraints obtained from chiral effective field theory~\cite{lynn2016,drischler2019}. Results for BSk20 and BSk21 are indistinguishable from those obtained for BSk26 and BSk24, respectively.}
\label{fig:NeuM-BSk}
\end{center}
\end{figure}

\begin{figure}[ht]
\begin{center}
\includegraphics[width=0.75\textwidth]{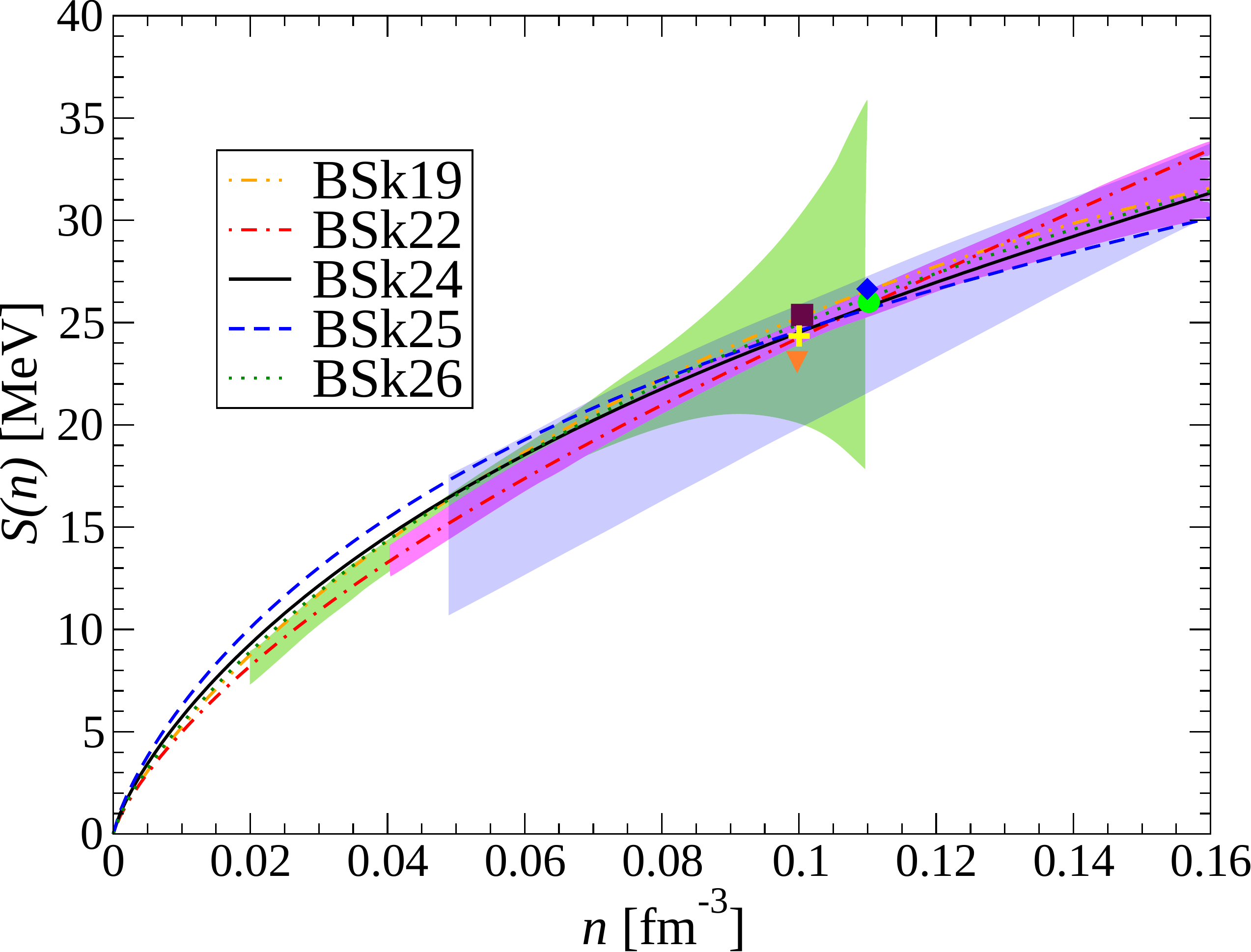}
\caption{(Color online) Variation of the symmetry energy $S(n)$ at subsaturation densities $n$ for the functionals BSk19, BSk20, BSk21, BSk22, BSk24, BSk25, and BSk26. The shaded areas are experimental constraints: from heavy-ion collisions~\cite{tsang2009} (blue), from isobaric-analog states and neutron skins~\cite{danielewicz2014} (purple), and from the electric dipole polarizability of $^{208}$Pb~\cite{zhang2015} (green). Symbols are values inferred from doubly magic nuclei~\cite{brown2013} (square), Fermi-energy difference~\cite{wang2013} (circle), binding-energy differences among heavy isotope pairs~\cite{zhang2013} (diamond), giant dipole resonance in $^{208}$Pb~\cite{trippa2008} (cross), giant quadrupole resonance in $^{208}$Pb~\cite{roca-maza2013} (triangle). Results for BSk20 and BSk21 are indistinguishable from those obtained for BSk26 and BSk24, respectively.
}
\label{fig:esym-exp}
\end{center}
\end{figure}

\begin{figure}[ht]
\begin{center}
\includegraphics[width=0.75\textwidth]{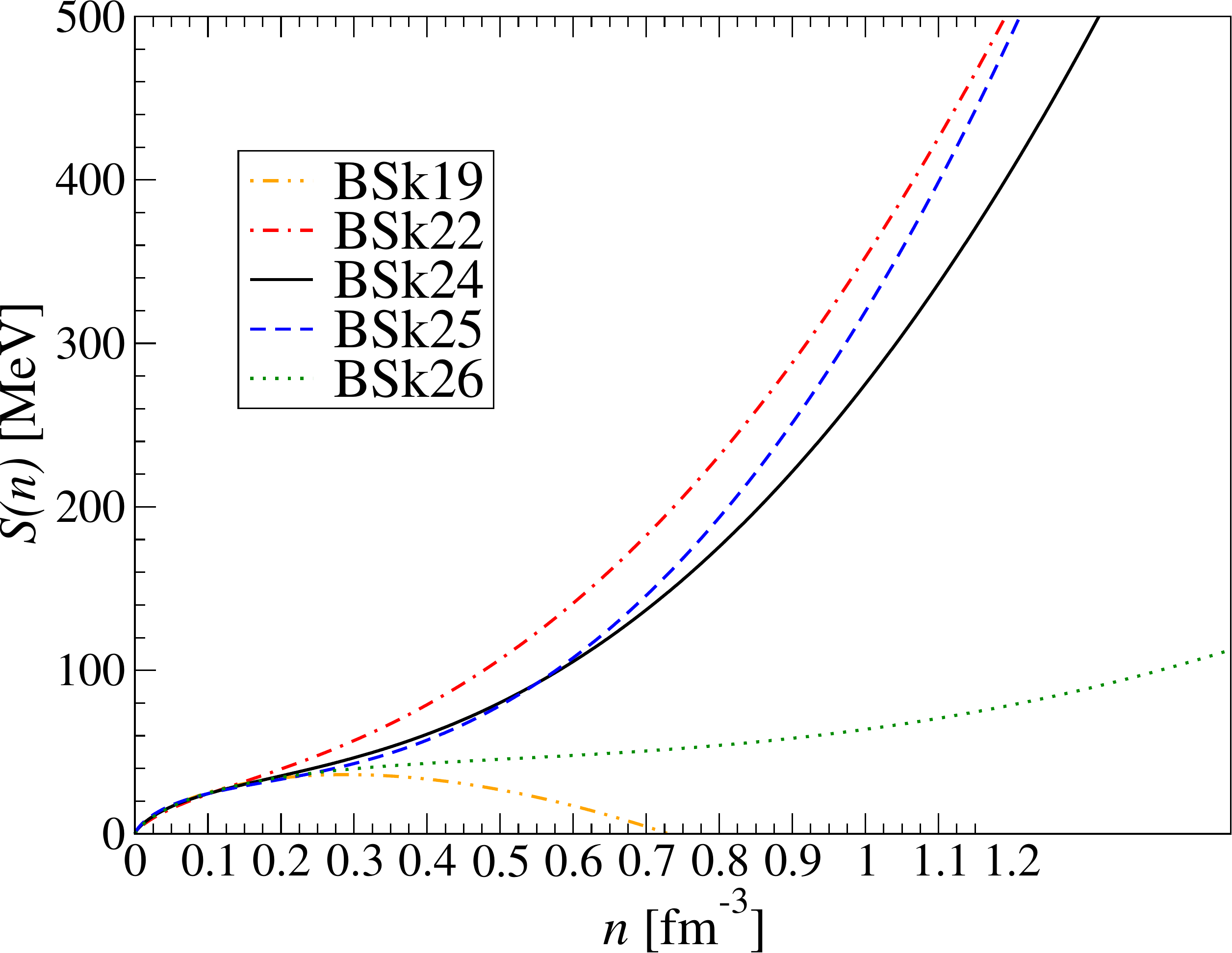}
\caption{(Color online) Variation of the symmetry energy $S(n)$ with density $n$ for the functionals BSk19, BSk20, BSk21, BSk22, BSk24, BSk25, and BSk26. Results for BSk20 and BSk21 are indistinguishable from those obtained for BSk26 and BSk24, respectively.}
\label{fig:esym-high}
\end{center}
\end{figure}

To ensure reliable extrapolations to the highly neutron-rich and very dense interiors of NSs,  the functionals were further constrained to reproduce the EoS of homogeneous NeuM, as calculated by many-body theory. Although the EoS is fairly well determined at densities below the saturation density $n_0=0.16$~fm$^{-3}$, it remains highly uncertain at supersaturation densities prevailing in the core of the most massive NSs. Two different EoSs were considered: the rather stiff EoS labeled as `V18' by Li and Schulze~ \cite{ls2008} 
and the softer EoS labeled as `A18 + $\delta\,v$ + UIX$^*$' by Akmal, Pandharipande and Ravenhall~\cite{apr1998}. 
The fit to nuclear masses along with these constraints does not lead to a unique determination of the functional. 
Expanding the energy per nucleon of infinite nuclear matter (INM) of density $n = n_0(1 + \epsilon)$  and charge asymmetry $\eta = (n_n - n_p)/n$ about the equilibrium density $n=n_0$ and $\eta = 0$, 
\begin{equation}
e(n, \eta) = a_v + \left(J + \frac{1}{3}L\epsilon\right)\eta^2 +
\frac{1}{18}(K_v + \eta^2K_\mathrm{sym})\epsilon^2 + \cdots 
\end{equation} 
the incompressibility coefficient $K_v$ was further restricted to lie in the experimental range $K_v=240\pm10$~MeV~\cite{colo2004}. 
To achieve a good fit to nuclear masses with a root-mean-square deviation as low as $0.5-0.6$ MeV, it was necessary to limit the allowed values of the symmetry-energy coefficient $J$ from 29 to 32 MeV. The parameters $L$ and $K_{\rm sym}$ were constrained by the fit to the NeuM EoS. The functionals BSk22, BSk23, BSk24 and BSk25 were all fitted to the NeuM EoS of Ref.~\cite{ls2008} while having $J$ = 32, 31, 30 and 29 MeV, respectively. To assess the role of the NeuM EoS, the functional BSk26 was fitted to the softer EoS of Ref.~\cite{apr1998} with $J=30$~MeV. Nuclear-matter properties for these functionals are summarized in Table~\ref{tab1}. The intermediate functional BSk23 will not be further considered here. As shown in Fig.~\ref{fig:NeuM-BSk}, these functionals are consistent with recent NeuM calculations based on chiral effective field theory~\cite{lynn2016,drischler2019}.  All functionals are also consistent with constraints on the EoS of SNM inferred from heavy-ion collisions~\cite{danielewicz2002,lynch2009}. As shown in Fig.~\ref{fig:esym-exp}, the variation of the symmetry energy $S(n)$ with density $n$ as predicted by the Brussels-Montreal functionals are compatible with experimental constraints from transport-model analyses of midperipheral heavy-ion collisions of Sn isotopes~\cite{tsang2009}, from the analyses of isobaric-analog states and neutron-skin data~\cite{danielewicz2014}, and from the electric dipole polarizability of $^{208}$Pb~\cite{zhang2015}. For comparison, we have also shown the following estimates: $S(0.1~\text{fm}^{-3})=25.5\pm1.0$~MeV from doubly magic nuclei~\cite{brown2013}, $S(0.11~\text{fm}^{-3})=26.2\pm1.0$~MeV from Fermi-energy difference~\cite{wang2013}, $S(0.11~\text{fm}^{-3})=26.65\pm0.20$~MeV from binding-energy differences among heavy isotope pairs~\cite{zhang2013}, $S(0.1~\text{fm}^{-3})=24.1\pm0.8$~MeV from giant dipole resonance in $^{208}$Pb~\cite{trippa2008},  $S(0.1~\text{fm}^{-3})=23.3\pm0.6$~MeV  from giant quadrupole resonance in $^{208}$Pb~\cite{roca-maza2013}. Note that different definitions were employed in these analyses. 
The symmetry energy is defined here as the difference between the energy per nucleon in NeuM and the energy per nucleon in SNM,  
 \begin{equation}
 S(n) = e_\mathrm{NeuM}(n) - e_\mathrm{SNM}(n)\, ,
\end{equation} 
where $e_\mathrm{NeuM}(n) \equiv e(n, 1)$ and $e_\mathrm{SNM}(n) \equiv e(n,0)$. Defining the symmetry energy as $(1/2)\, \partial^2e/\partial\eta^2$ (calculated for $\eta=0$)  leads to slightly different results (see, e.g., Refs.~\cite{gcp2010,gcp2013} for discussions); the deviations amount to about 1 MeV at most for the densities shown in Fig.~\ref{fig:esym-exp}, and are therefore much smaller than the current overall experimental uncertainties. The functionals mainly differ in their predictions for the symmetry energy at densities $n>n_0$, as shown in Fig.~\ref{fig:esym-high}. 

\begin{table}[ht]
\caption{Nuclear-matter properties for the Brussels-Montreal functionals. The last line indicates the NeuM EoS to which each functional was fitted: FP~\cite{Friedman1981}, APR~\cite{apr1998}, and LS2~\cite{ls2008}.}
\label{tab1}
\vspace*{0.2 cm}
\begin{tabular}{|c|ccc|ccccc|}\hline
& BSk19&BSk20&BSk21&BSk22&BSk23&BSk24&BSk25&BSk26\\
\hline
$J$ {[MeV]} &30.0&30.0&30.0&32.0&31.0&30.0&29.0&30.0\\
$L$ {[MeV]} &31.9&37.4&46.6&68.5&57.8 &46.4&36.9&37.5\\
$K_v$ {[MeV]} &237.3&241.4&245.8&245.9&245.7 &245.5 & 236.0 & 240.8\\
$K_\mathrm{sym}$ {[MeV]} &-191.4&-136.5&-37.2&13.0& -11.3 & -37.6 & -28.5 & -135.6\\
NeuM & FP & APR & LS2 & LS2 & LS2 & LS2 & LS2 & APR \\
\hline
\end{tabular}
\end{table}

To better assess the role of the NeuM EoS, we will also consider the older series BSk19, BSk20 and BSk21~\cite{gcp2010} for which unified EoSs for NSs have been also calculated~\cite{pearson2011,pcgd2012}. These functionals were fitted to older atomic mass data from the 2003 Atomic Mass Evaluation~\cite{AME2003} with the same symmetry-energy coefficient $J=30$~MeV. While BSk20 (BSk21) was simultaneously adjusted to the same NeuM EoS as BSk26 (BSk24), BSk19 was constrained to reproduce the softer NeuM EoS of Ref.~\cite{Friedman1981}. Although the unified EoS corresponding to BSk19 fails to explain the existence of the most massive NSs~\cite{chamel2011}, it may still be applicable to the medium-mass NSs observed by the LIGO-Virgo collaboration. As a matter of fact, functional BSk19 still remains compatible with recent \textit{ab initio}  calculations~\cite{lynn2016,drischler2019}, see Fig.~\ref{fig:NeuM-BSk}. These older functionals are also consistent with experimental constraints on the symmetry energy, as shown in Fig.~\ref{fig:esym-exp}. The BSk19 functional leads to a very soft symmetry energy at higher densities, as can be seen in Fig.~\ref{fig:esym-high}. Nuclear-matter parameters are summarized in Table~\ref{tab1}.

\subsection{Consistent description of the different regions of a neutron star}
\label{sec:model}

Below its thin atmosphere and liquid ``ocean'', a NS is thought to contain at least three distinct regions: an outer crust made of neutron-rich nuclei in a charge neutralizing electron background, an inner crust consisting of neutron-proton clusters immersed in a neutron sea (possibly enriched with protons at sufficiently high densities), and a liquid core of nucleons and leptons. Other particles such as hyperons might exist in the central core of most massive NSs, but we will not consider this possibility here. 

A detailed account of the calculations of the EoS in the different regions of a NS can be found in Ref.~\cite{pearson2018}. We recall only the main features here. The equilibrium properties of the outer crust for densities $\rho \gtrsim 10^6$~g~cm$^{-3}$ were determined by minimizing the Gibbs free energy per nucleon $g$ at each given pressure $P$ assuming pure layers with a perfect body-centered cubic crystal structure~\cite{pearson2011}. The EoS in this region was calculated making use of experimental data supplemented by HFB mass tables for nuclei whose mass has not been measured. At the pressure $P_{\rm drip}$ such that $g=M_n c^2$,  neutrons drip out of nuclei marking the transition to the inner crust~\cite{cfzh2015}. Full HFB calculations beyond this point would be computationally extremely costly due to the widely different scales involved. For this reason, the fourth-order extended Thomas-Fermi method was adopted within the Wigner-Seitz approximation using parametrized nucleon density distributions. Proton shell and pairing corrections were added perturbatively via the Strutinsky integral  theorem~\cite{pcgd2012,pcpg2015}. At densities high enough for free protons to appear, the shell and pairing corrections were dropped entirely. 
For convenience, the energy per nucleon was minimized at fixed  average baryon number density. As shown in Ref.~\cite{pcgd2012}, density discontinuities are negligibly small in the inner crust so that minimizing the energy per nucleon or the Gibbs free energy per nucleon yields practically the same results without having recourse to a Maxwell construction. The pressure $P$ at any mean density $\bar{n}$ was calculated semi-analytically as described in Appendix B of Ref.~\cite{pcgd2012}. Calculations in the inner crust were performed using the same functional as that underlying the HFB nuclear mass model used in the outer crust. Calculations in the core were comparatively much simpler since the energy density and the pressure obtained from the energy-density functional theory are given by analytic expressions (see Ref.~\cite{pearson2018}). Complete numerical results and analytic fits applicable to the entire star can be found in Refs.~\cite{potekhin2013,pearson2018}. 

\section{Tidal deformability of a neutron star}
\label{sec:tidal}

The formalism to calculate the structure and the tidal deformability of a NS is reviewed in Section~\ref{sec:tidal-formalism}. The role of the symmetry energy and NeuM is studied in Section~\ref{sec:tidal-micro}. 

\subsection{Calculation of the Love number and tidal deformability}
\label{sec:tidal-formalism}

We summarize here the main equations that are needed to compute the second gravito-electric tidal Love number $k_2$. More details can be found, e.g., in Refs.~\cite{flanagan08, hinderer08, hinderer10}.

Let us consider a star that is both static and spherically symmetric. Once placed in a static external quadrupolar tidal field $\mathcal{E}_{ij}$ (e.g., associated with the gravitational field of a companion for a NS in a binary system), this star will acquire a non-zero quadrupole moment $Q_{ij}$ whose expression, to linear order, will simply read
\begin{equation}
Q_{ij} = -\lambda\,  \mathcal{E}_{ij}\, .
\end{equation}
The quantity $\lambda$ characterizes the response of the star (through its induced quadrupole moment $Q_{ij}$) to a given applied quadrupolar tidal field $\mathcal{E}_{ij}$. It is related to the dimensionless $\ell=2$  tidal Love number $k_2$ through  
\begin{gather}
k_2 = \frac{3}{2}\, G\,  \lambda\, R^{-5}\, .
\end{gather}  
Note that both $\lambda$ and $k_2$ depend on the structure of the star and therefore on the mass and the EoS of dense matter. Denoting by $C$ the star's compactness parameter, i.e.,
\begin{equation}
    C = \dfrac{G\, M}{R\, c^2}\, ,
\end{equation}
the Love number $k_2$ can be shown to be expressible as~\cite{hinderer08}
\begin{align}
k_2 =& \,\frac{8C^5}{5}(1-2C)^2\big[2+2C(y-1)-y\big] \, \Big\{2C\big[6-3y+3C(5y-8)\big] \nonumber \\ &+ 4C^3\big[13-11y+C(3y-2)+2C^2(1+y)\big] \nonumber \\ &+ 3(1-2C)^2\big[2-y+2C(y-1)\big] \ln (1-2C) \Big\}^{-1} \, .
\end{align}
The quantity $y\equiv R\, H'(R)/H(R)$ involved in this last equation can be obtained by integrating the following differential equation
\begin{align}
\label{H_eq}
&H''(r)+ H'(r) \bigg(1-\frac{2 G m(r)}{c^2r}\bigg)^{-1} \Bigg\{\frac{2}{r} - \frac{2G m(r)}{c^2r^2} - \frac{4 \pi G}{c^4}\, r \big(\mathcal{E}(r) - P(r)\big) \Bigg\} \nonumber\\& + H(r) \bigg(1-\frac{2G m(r)}{c^2r}\bigg)^{-1} \Bigg\{\frac{4\pi G}{c^4}\bigg[5\mathcal{E}(r) + 9P(r) +\dfrac{d\mathcal{E}}{dP}(r) \big(\mathcal{E}(r)  + P(r)\big)\bigg] \nonumber  \\& - \frac{6}{r^2} - 4\bigg(1-\frac{2Gm(r)}{c^2r}\bigg)^{-1} \bigg(\frac{Gm(r)}{c^2r^2} + \frac{4\pi G}{c^4}\, r\,  P(r)\bigg)^2 \Bigg\} = 0 \, ,
\end{align}
 where $\mathcal{E}(r)$ and $P(r)$ denote respectively the mass-energy density and the pressure of matter at the radial (circumferential) coordinate $r$ and $m(r)$ is the mass enclosed in a circular contour of radius $r$.  Once an EoS has been prescribed (in the form $P=P\left(\mathcal{E}\right)$), Eq.~\eqref{H_eq} can be integrated together with the Tolman-Oppenheimer-Volkoff (TOV) equations~\cite{tolman1939,oppenheimer1939}
\begin{equation}
\frac{{\rm d}P(r)}{{\rm d}r} = -\frac{G\, \mathcal{E}(r)m(r)}{c^2 r^2}
\biggl[1+\frac{P(r)}{\mathcal{E}(r)}\biggr] \biggl[1+\frac{4\pi P(r)r^3}{c^2m(r)}\biggr]\biggl[1-\frac{2Gm(r)}{c^2 r}\biggr]^{-1}\, ,
\end{equation}
and 
\begin{equation}
m(r) = \frac{4\pi}{c^2}\int_0^r\mathcal{E}(r')r'^{\, 2}\, {\rm d}r'\, ,
\end{equation}
with the boundary conditions 
\begin{equation}
    m\left(0\right)=0\, , \ \ \ \ \mathcal{E}\left(0\right) = \mathcal{E}_c\, , \ \ \ \ H(0) = 0  \ \ \ \ \text{and} \ \ \ \ H'(0) = 0 \, ,
\end{equation}
where $\mathcal{E}_c$ is the mass-density at the center of the star. The gravitational mass of the star is thus given by $M=m(R)$ where $R$ is the radial coordinate at which the radiative surface is reached, i.e. $P(R) = 0$. In what follows, this set of equations is numerically solved by means of a fourth-order Runge-Kutta method, using the analytical fits of the EoSs presented in  Refs.~\cite{potekhin2013,pearson2018}. 

\subsection{Dependence of the tidal coefficients on the symmetry energy and on NeuM}
\label{sec:tidal-micro}

\subsubsection{Relations between $\Lambda_{1.4}$ and $R_{1.4}$}
\label{sec:rel_Lambda_R}

\begin{figure}[ht]
\begin{center}
\includegraphics[width=0.85\textwidth]{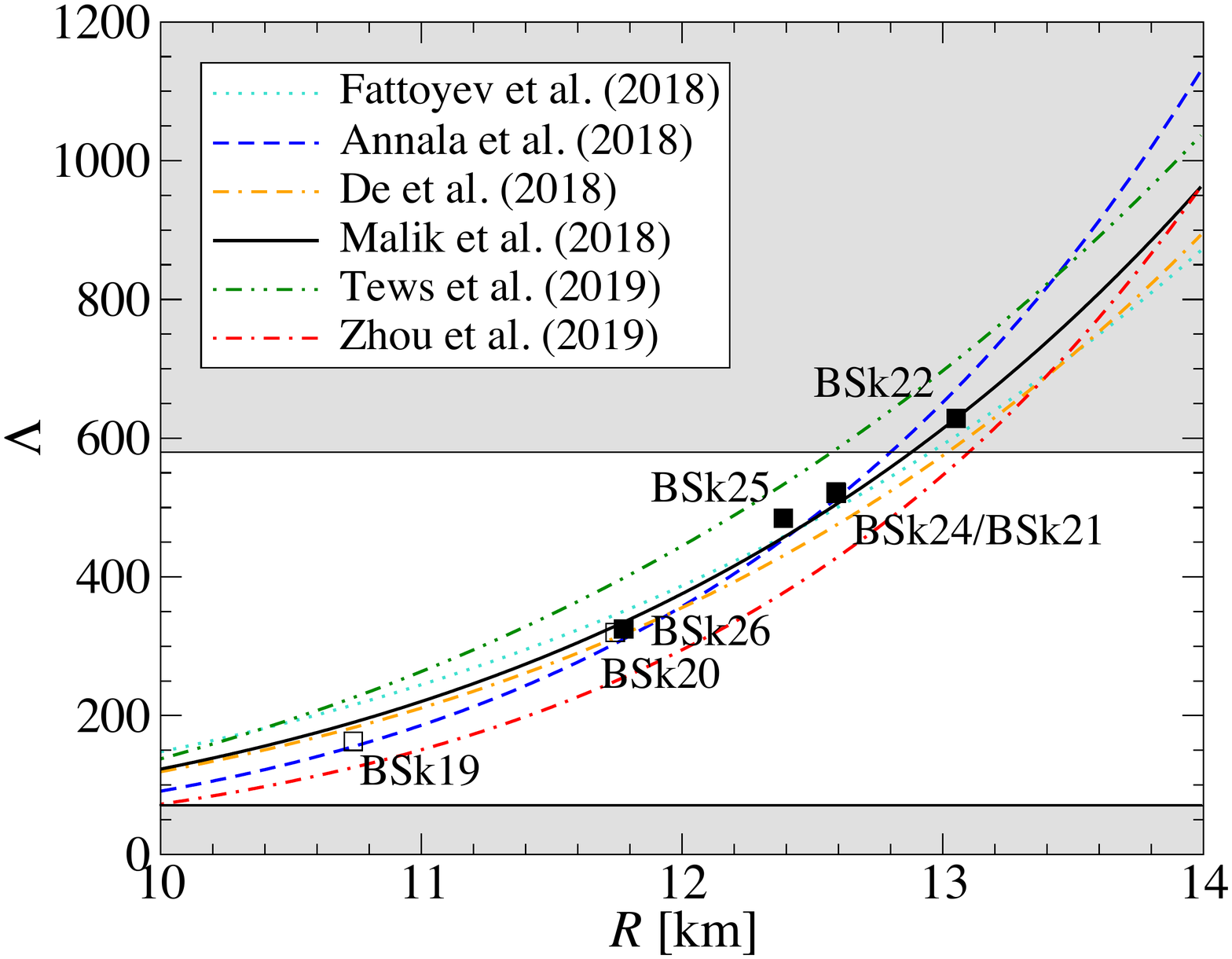}
\caption{(Color online) Tidal deformability coefficient $\Lambda$ of a $1.4~M_\odot$ NS as a function of the radius $R$, as predicted by selected empirical relations proposed by various authors (lines). For comparison, results obtained using the unified EoSs considered in this work are also shown (symbols). The central area represents the LIGO-Virgo constraint $\Lambda = 190^{+390}_{-120}$ at 90\% confidence level using method (ii)~\cite{ligo2018}. See text for details.}
\label{fig:lambda14}
\end{center}
\end{figure}

The tidal deformability coefficient $\Lambda_{1.4}$ of a $1.4~M_\odot$ NS has been shown to be strongly correlated with the corresponding NS radius $R_{1.4}$. However, different empirical relations of the form $\Lambda_{1.4}\propto R_{1.4}^{\, \alpha}$ have been proposed with $\alpha$ ranging from $5$ to $7.71$ ~\cite{fattoyev2018,de2018,annala2018,malik2018,zhou19,lourenco2019a,lourenco2019b,tong2019,nandi19constraining,tews19,tsang2019b} (the relation proposed in Ref.~\cite{tews19} includes a constant shift). As shown in Fig.~\ref{fig:lambda14}, most of these relations equally well reproduce the results we obtained with the unified Brussels-Montreal EoSs with deviations of a few \%, except for those proposed in Refs.~\cite{zhou19,tews19} for which the deviations amount to $10-20$~\%. Comparing BSk22, BSk24, and BSk25, which were constrained to the same NeuM EoS, confirms that $\Lambda_{1.4}$ depends on the symmetry energy. Comparing BSk24 and BSk26, which were both fitted with the same value for $J=30$~MeV but different NeuM EoSs therefore different values for the slope $L$ of the symmetry energy (see Table~\ref{tab1}), shows that $\Lambda_{1.4}$ increases with $L$, as found in previous studies (e.g., in Ref.~\cite{fattoyev2018}).

\subsubsection{Role of the symmetry energy and of NeuM on the tidal deformability}

\begin{figure}[ht]
\begin{center}
\includegraphics[width=0.85\textwidth]{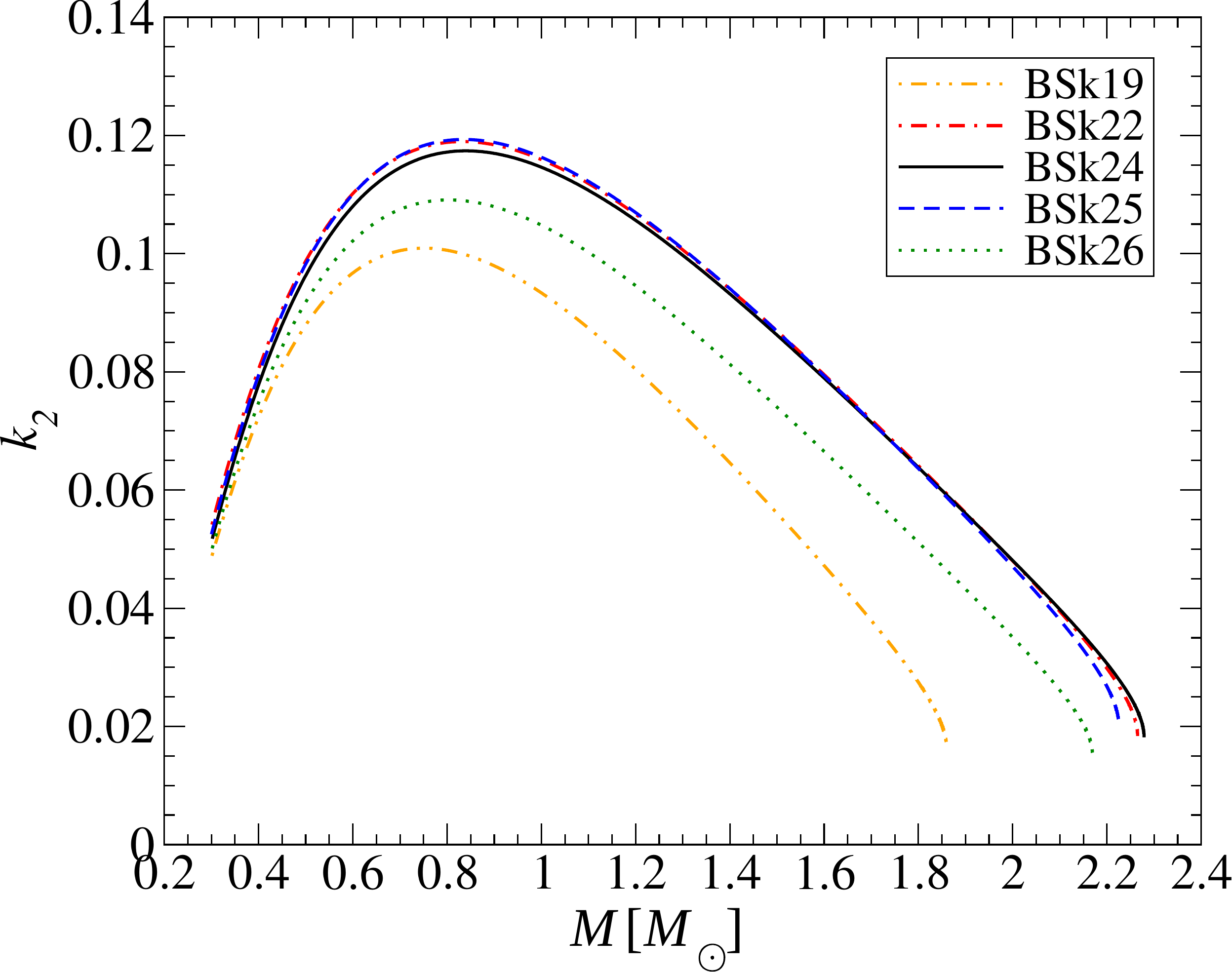}
\caption{(Color online) Second gravito-electric Love number $k_2$ as a function of NS mass, as predicted by the unified EoSs of Refs.~\cite{potekhin2013,pearson2018}. See text for details.}
\label{fig:k2-BSk}
\end{center}
\end{figure}

The dependence of $\Lambda$ on the symmetry energy comes to a large extent from the factor $R^5$ (see Eq.~\eqref{eq:Lambda}). Indeed, comparing BSk22, BSk24, and BSk25 shows that $k_2$ is essentially independent of the symmetry energy, as can be seen in Fig.~\ref{fig:k2-BSk}. On the contrary, the role of the symmetry energy on NS radii is well-known (see, e.g., Refs.~\cite{fortin2016,margueron2018}) and is also apparent in the predictions from the EoSs considered in this work, as previously discussed in Ref.~\cite{pearson2018}. Malik \textit{et al}.~\cite{malik2018} have recently examined possible correlations between tidal deformabilities and nuclear-matter parameters for a large set of EoSs. As can be seen from their results in Table I for a few selected NS masses, the symmetry-energy coefficient $J$ (denoted by $J_0$ in their paper) has essentially no impact on $k_2$ and the slope $L$ of the symmetry energy at saturation (denoted by $L_0$ in their paper) has only a moderate influence on $k_2$ (a similar conclusion can be drawn from the recent analysis of Ref.~\cite{raithel2019b} considering a large set of parametrized polytropic EoSs). This suggests that the minor role played by the symmetry energy on $k_2$ is not a conclusion restricted to the EoSs adopted in this work but is actually quite robust. However, it should be stressed that unlike the EoSs considered here, those selected in Ref.~\cite{malik2018} were based on widely different functionals that were constructed following different fitting protocols. In particular, those functionals differ in their predictions not only for the symmetry energy but also for other nuclear properties. 

The comparison between BSk24 and BSk26, also shown in Fig.~\ref{fig:k2-BSk}, reveals that $k_2$ is more sensitive to the stiffness of the NeuM EoS. This dependence is more apparent on the older series BSk19, BSk20, and BSk21, whose results for $k_2$ are displayed in Fig.~\ref{fig:k2-BSk}. These EoSs were fitted to three different NeuM EoSs with very different degrees of stiffness: BSk19 corresponding to the softest and BSk21 to the stiffest. Results for BSk20 and BSk21 are indistinguishable from those obtained for BSk26 and BSk24 respectively. As shown in Fig.~\ref{fig:k2-BSk}, the NeuM EoS is found to have essentially no effect on $k_2$ for NSs with a mass $M\lesssim 0.5~M_\odot$. This merely stems from the fact that the NeuM EoS is very tightly constrained by \textit{ab initio} calculations at densities below about twice saturation density. The large uncertainties on the NeuM EoS at higher densities has a strong influence on $k_2$ for NSs with a mass $M>0.5~M_\odot$. The impact of the NeuM EoS on $k_2$ is the strongest for the most massive NSs. The stiffer the NeuM EoS is, the larger is $k_2$.

\section{Analysis of GW170817}
\label{sec:GW170817}

General constraints on the structure of nonrotating NSs, as inferred from analyses of gravitational-wave data from GW170817 during the inspiral phase and from observations of electromagnetic counterparts, are discussed in Section~\ref{sec:gw170817-general}. Constraints obtained from measurements of the tidal deformability are discussed in Section~\ref{sec:gw170817-tidal}.
 
\subsection{General constraints on the structure of nonrotating neutron stars}
\label{sec:gw170817-general}

\subsubsection{Constraints on the radius}

According to numerical simulations, the amount of material ejected during the collision of the NSs can be traced back to the fate of the compact remnant. The rather large estimates $\sim 0.02-0.05~M_\odot$ inferred from observations of the electromagnetic counterpart of GW170817 (see, e.g., Ref.~\cite{cote2018} for a compilation)  point against a prompt collapse to a black hole. If this scenario is correct, numerical simulations show that the total mass $M_{\rm tot}=M_1+M_2$ of the two NSs must be lower than some threshold value $M_{\rm thres}$. Using the measured value $M_{\rm tot}=2.74^{+0.04}_{-0.01}~M_\odot$ from gravitational-wave observations~\cite{ligo2017inspiral} together with an empirical relation for $M_{\rm thres}$ and the causality condition, Bauswein \textit{et al}.~\cite{bauswein2017} obtained the following lower limit on the radius of a $1.6~M_\odot$ NS: $R_{1.6} \geq 10.30^{+0.15}_{-0.03}$ km. 
Assuming further that the remnant lived for more than 10 ms, they obtained the more stringent constraint $R_{1.6} \geq 10.68^{+0.15}_{-0.04}$ km. 
Note that the estimated uncertainties in these constraints do not take into account the systematic errors in the empirical relations. 
Using a different empirical relation for $M_{\rm thres}$ but similar arguments, the authors of Ref.~\cite{koppel19} derived a tighter bound on NS radii: $R\geq -0.88M^2+2.66M+8.91 $ km for $1.2~M_\odot <M<2~M_\odot$ \cite{koppel19}. 
As shown in Fig.~\ref{fig:MR-BSk}, these constraints are fulfilled by all seven EoSs considered in this work except for BSk19. The exclusion of BSk19 should come as no surprise since the empirical relations for $M_{\rm thres}$ were obtained by selecting EoSs that are consistent with the existence of massive NSs. 

\subsubsection{Constraints on the maximum mass}

\begin{figure}[!t]
\centering
\includegraphics[width=0.85\linewidth]{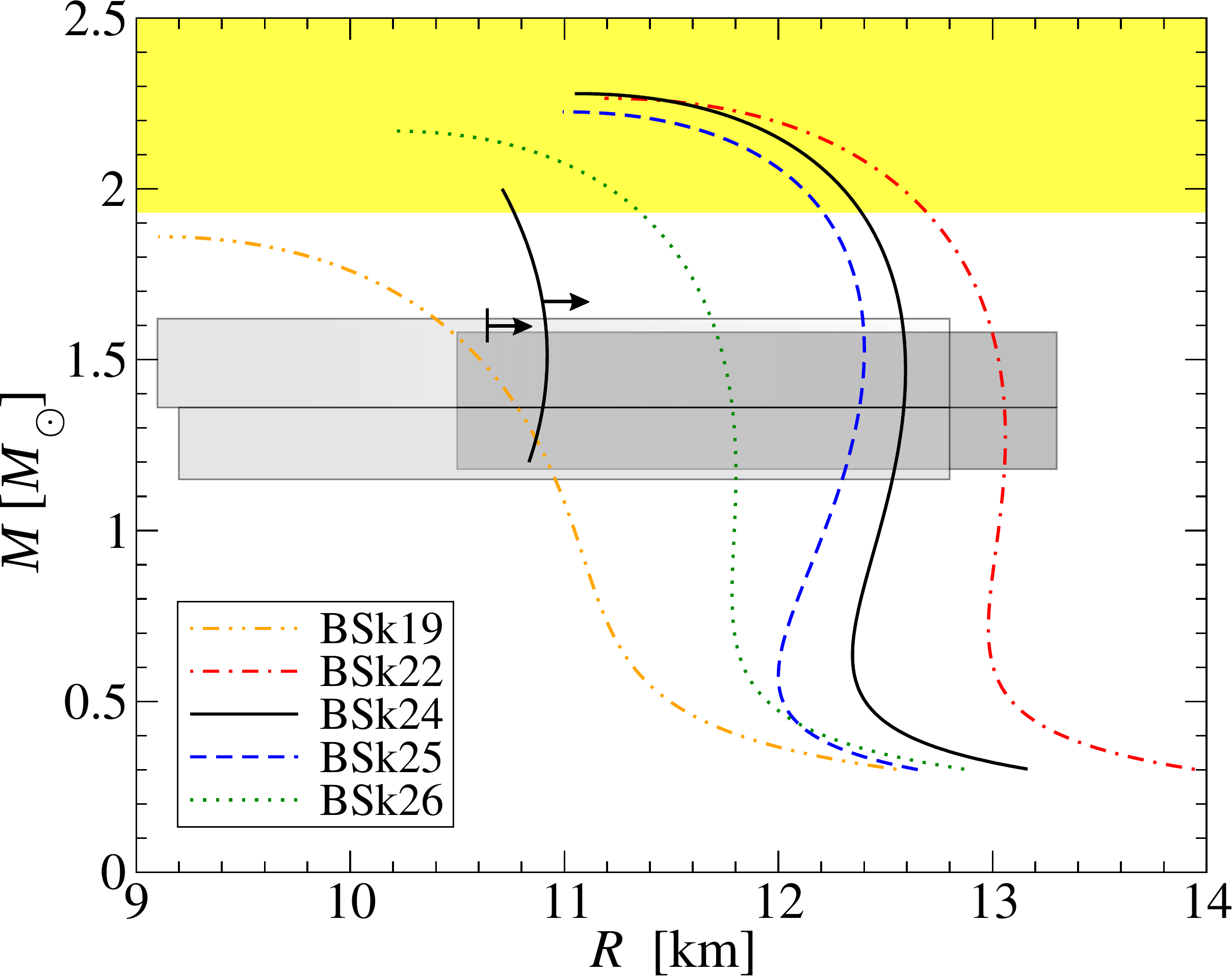}
\caption{(Color online) Gravitational mass $M$ (in solar masses) versus circumferential radius $R$ of nonrotating NSs, calculated using the unified EoSs of Refs.~\cite{potekhin2013,pearson2018}. Results for BSk20 and BSk21 are indistinguishable from those obtained for BSk26 and BSk24 respectively. Constraints inferred from the analysis of GW170817 are also shown: lines with arrows are lower bounds on the radius from Refs.~\cite{bauswein2017,koppel19}; dark and light boxes are the masses and radii of the two NSs (90\% confidence level) as inferred by the LIGO-Virgo collaboration using method (iii) with the additional requirement $M_\text{max} \geq 1.97~M_\odot$, and (ii) respectively; the yellow area denotes expected values for the maximum mass as obtained by various studies. See text for details.}
\label{fig:MR-BSk}
\end{figure}

Different analyses of the short gamma-ray burst and of the kilonova emission, combined with the total binary mass $M_{\rm tot}=2.74^{+0.04}_{-0.01}~M_\odot$ inferred from gravitational-wave observations, have led to constraints on the maximum  mass of a nonrotating NS. Assuming the formation of a short-lived NS, Margalit and Metzger~\cite{margalit17} obtained $M_\text{max} \lesssim 2.17~M_\odot$ (90\% confidence), Rezzolla \textit{et al}.~\cite{rezzolla18} $M_\text{max} \lesssim 2.16^{+0.17}_{-0.15}~M_\odot$ (90\% confidence), and Ruiz \textit{et al}.~\cite{ruiz18}  $2.16 \pm 0.23~M_\odot \lesssim M_\text{max}  \lesssim 2.28\pm 0.23~M_\odot$. Shibata \textit{et al}.~\cite{shibata17,shibata19} obtained compatible estimates, $2.1~M_\odot \lesssim M_\text{max}  \lesssim 2.3~M_\odot$, under the assumption of a longer-lived NS (with a lifetime up to tens of seconds). Combining all studies, conservative limits on the maximum mass are $1.93~M_\odot \lesssim M_\text{max}  \lesssim 2.51~M_\odot$. As shown in Fig.~\ref{fig:MR-BSk}, all seven considered EoSs but BSk19 are consistent with these constraints. Alternatively, other authors have interpreted the late-time electromagnetic emission in terms of a very long-lived NS remnant (with a lifetime of about 20 days)~\cite{ai2018, geng2018, li2018, yu2018} and concluded that $M_\text{max} \gtrsim 2.6~M_\odot$~\cite{yu2018}. If this latter scenario is correct, all EoSs would be ruled out. 

Any firm conclusion on the EoS can hardly be drawn in view of the lack of consensus on the interpretation of the electromagnetic counterparts of GW170817. 

\subsection{Constraints on the tidal deformability of a neutron star}
\label{sec:gw170817-tidal}

\begin{figure}[!t]
\centering
\includegraphics[width=0.85\linewidth]{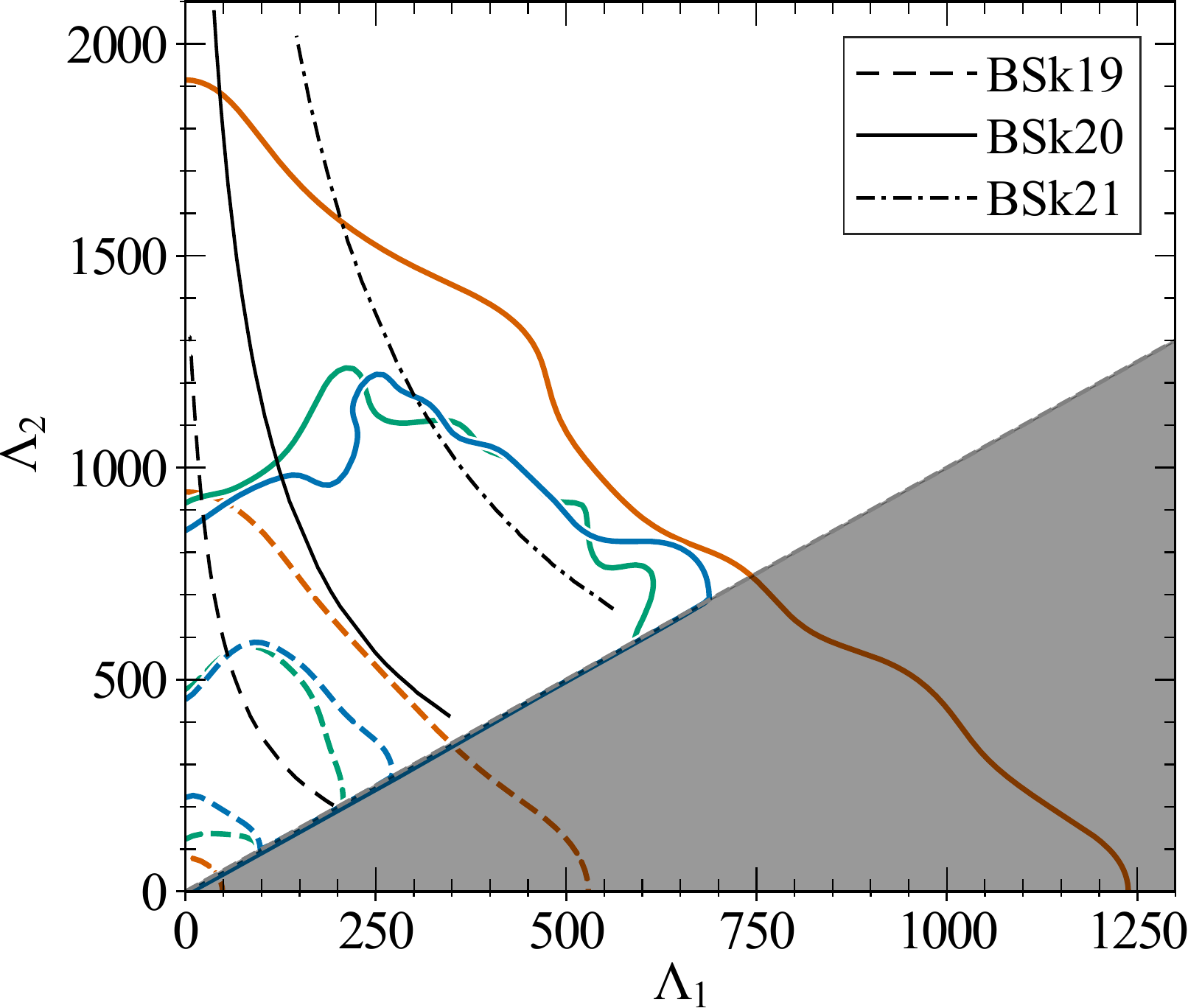}
\caption{(Color online) Dimensionless tidal deformability parameter of the secondary star as a function of that of the primary star ($M_1>M_2$), as predicted by the unified EoSs of Ref.~\cite{potekhin2013} for a chirp mass $\mathcal{M} = 1.186~M_\odot$~\cite{ligo2019prx} (black lines). The gray shading corresponds to the unphysical region $\Lambda_2<\Lambda_1$. Colored curves are taken from Ref.~\cite{ligo2018}: the green, blue and orange lines denote 50~\% (dashed) and 90~\% (solid) credible levels for the posteriors obtained using EoS-insensitive relations (method ii), parametrized EoSs without any maximum-mass requirement (method iii) and independent EoSs (method i). See text for details.}
\label{fig:lambda12-BSk19-21}
\end{figure}

\begin{figure}[!t]
\centering
\includegraphics[width=0.85\linewidth]{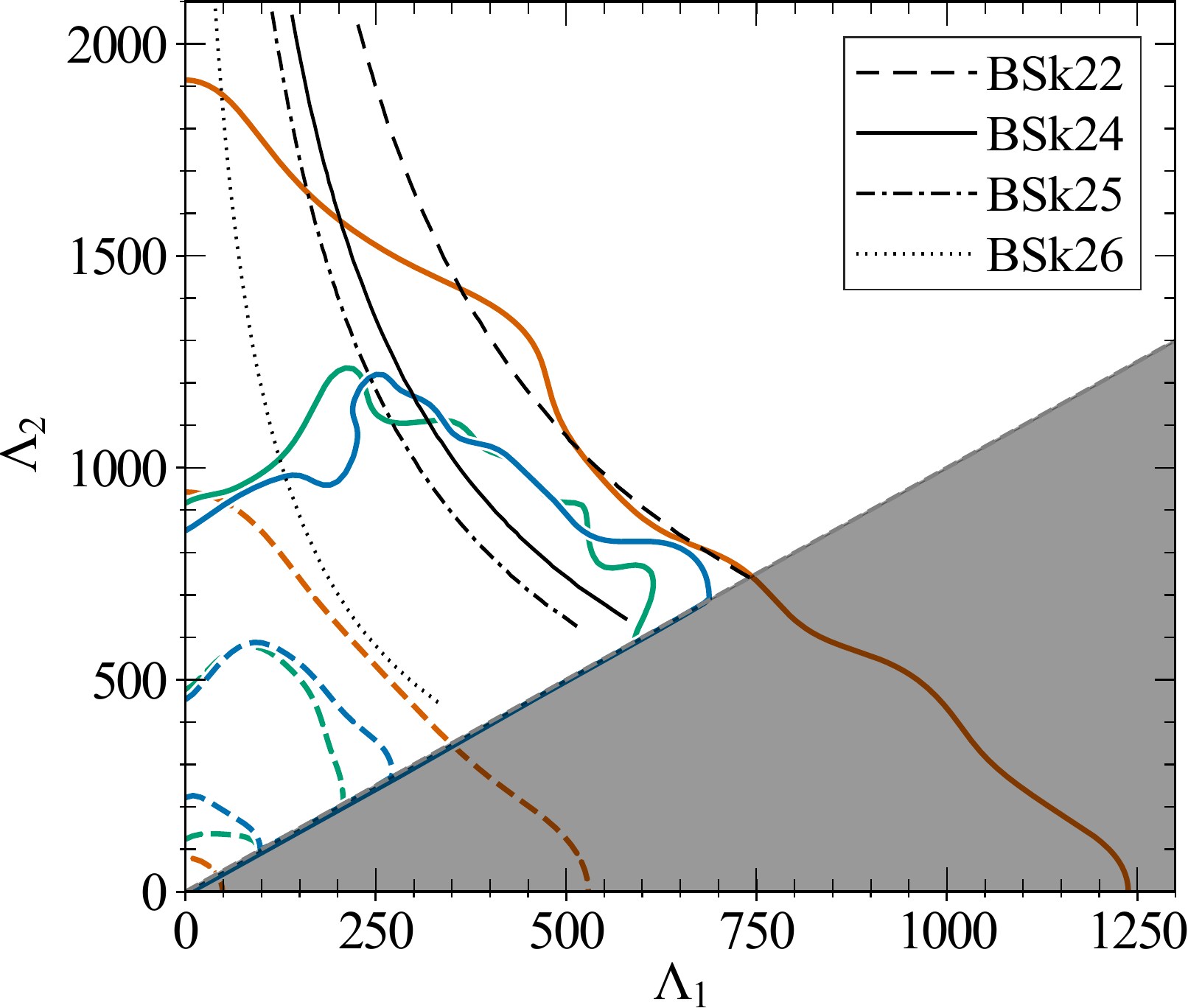}
\caption{(Color online) Same as Fig.~\ref{fig:lambda12-BSk19-21} for the unified EoSs of Ref.~\cite{pearson2018}.}
\label{fig:lambda12-BSk22-26}
\end{figure}

The latest analysis of the full gravitational-wave signal from the LIGO-Virgo collaboration~\cite{ligo2018} has led to constraints on the tidal deformability parameters of the two NSs by comparing the gravitational-wave data to theoretical waveforms 
using low-spin priors as expected from galactic binary NS spin measurements. Three different methods were employed: (i) the NS masses $M_1$, $M_2$ and the tidal deformabilities $\Lambda_1$, $\Lambda_2$ were treated independently, (ii) $\Lambda_1$, $\Lambda_2$, and $M_2/M_1$ were related by a universal (i.e., EoS-insensitive) relation (implying that the two NSs are described by the same EoS), (iii) a large set of parametrized EoSs was used ensuring causality (assuming a common EoS for the two NSs). As can be seen in Figs.~\ref{fig:lambda12-BSk19-21} and \ref{fig:lambda12-BSk22-26}, all EoSs but BSk22 are consistent with the inferred tidal deformabilities at the 90~\% credible level for all three methods. The BSk22 EoS is only marginally compatible with the 90~\% credible level for method (iii) and ruled out by the first two methods. Interestingly, among the seven EoSs considered here, BSk19 is the only one that lies within the 50~\% credible levels thus confirming that the gravitational-wave data alone tend to favor a rather soft EoS at densities relevant for medium-mass NSs. Incidentally, the analyses of K$^+$ production~\cite{fuchs2001,sturm2001,hartnack2006} and $\pi^-/\pi^+$ production ratio~\cite{xiao2009} in heavy-ion collisions provide evidence for a soft EoS at similar densities. 

Combining empirical relations between $\Lambda_{1.4}$ and $R_{1.4}$ (see Section~\ref{sec:rel_Lambda_R}) with the upper limits on the tidal deformability obtained from the analyses of the gravitational-wave signal by the LIGO-Virgo collaboration~\cite{ligo2017inspiral, ligo2018} and De \textit{et al}.~\cite{de2018}, different constraints have been proposed for the radius of a $1.4~M_\odot$ NS (see, e.g., Refs.~\cite{raithel18,fattoyev2018,nandi18hybrid,malik2018,tong2019,zhou19}). 
Still, given the absence of an exact relation between the two quantities, the latest upper limit $\Lambda_{1.4}<580$ from the LIGO-Virgo collaboration using method (ii)~\cite{ligo2018} cannot yield a constraint on $R_{1.4}$ more accurate than $R_{1.4} \lesssim 12.6-13.1$~km, as can be seen from Fig.~\ref{fig:lambda14}. Incidentally, this figure confirms that all EoSs considered in this work but BSk22 are consistent with the inferred tidal deformability from the LIGO-Virgo collaboration. Less stringent constraints on $R_{1.4}$ were previously derived using parametrized EoSs and the initial upper estimate for $\Lambda_{1.4}$ from the LIGO-Virgo collaboration~\cite{ligo2017inspiral}, see, e.g., Refs.~\cite{annala2018,most18}. Considering more recent studies of the LIGO-Virgo data~\cite{tews19,radice19multi,coughlin18}, a conservative upper limit on the radius is $R_{1.4}<13.6$ km. Besides, we note that some of the studies previously mentioned have also deduced a rather tight constraint on the smallest possible radius of a $1.4~M_\odot$ NS using the lower limit on the tidal deformability found by Radice \textit{et al}.~\cite{radice18} from the analysis of the kilonova emission. Nevertheless, the results of Ref.~\cite{radice18} have been recently questioned~\cite{kiuchi19}. 

Actually, the LIGO-Virgo collaboration placed constraints on the $M-R$ diagram from the direct analysis of the gravitational-wave signal~\cite{ligo2018} (see also Ref.~\cite{de2018}). Using method (ii), the radii of the two NSs were thus estimated as $R_1=10.8^{+2.0}_{-1.7}$ km and $R_2=10.7^{+2.1}_{-1.5}$ km (at 90~\% confidence level) with $1.36~M_\odot \leq M_1\leq 1.62~M_\odot$ and $1.15~M_\odot \leq M_2\leq 1.36~M_\odot$ (at 90~\% level). As shown in Figs.~\ref{fig:MR-BSk}, all EoSs but BSk22 are compatible with these estimated masses and radii. Method (iii) with the additional requirement $M_\text{max} \geq 1.97~M_\odot$ (coming from pulsar observations~\cite{antoniadis13}) yielded $R_1=11.9^{+1.4}_{-1.4}$ km and $R_2=11.9^{+1.4}_{-1.4}$ km (at 90~\% confidence level) with $1.36~M_\odot \leq M_1\leq 1.58~M_\odot$ and $1.18~M_\odot \leq M_2\leq 1.36~M_\odot$ (at 90~\% level). 

\vspace*{0.4 cm}

\section{Conclusion}
\label{sec:conclusion}

The  role  of  the  symmetry  energy  and  the  NeuM  stiffness  on  the  tidal  deformability of a cold nonaccreted NS has been studied using the series of seven unified EoSs BSk19, BSk20, BSk21, BSk22, BSk24, BSk25, and BSk26 all based on the nuclear energy-density functional theory. These EoSs provide a thermodynamically consistent description of all regions of the stellar interior. The underlying functionals were precision fitted to various experimental and theoretical nuclear data. 

For the EoSs adopted in this work, the symmetry energy is found to have essentially no impact on the tidal Love number $k_2$. This implies that the tidal deformability parameter $\Lambda$ depends on the symmetry energy only through the radius $R$. The different predictions for $k_2$ mainly arise from uncertainties in the NeuM EoS at high densities. Since the radius $R$ of a NS also depends on the stiffness of the NeuM EoS, it is difficult to extract information on the symmetry energy and/or the NeuM EoS from the sole tidal deformability parameter~$\Lambda$. Still, the BSk22 EoS with a symmetry energy coefficient $J=32$ MeV and a slope $L=68.5$ MeV appears to be disfavored by the tidal-deformability estimates obtained by the LIGO-Virgo collaboration from the analysis of the gravitational-wave signal GW170817. Similarly, the analysis of the gravitational-wave signal alone is not very constraining for the NeuM EoS. In particular, predictions from the BSk19 EoS are consistent with the LIGO-Virgo tidal-deformability constraints even though this EoS does not support $2~M_\odot$ NSs. The gravitational-wave data alone thus tend to favor a rather soft EoS at densities relevant for medium-mass NSs, as also suggested by the analyses of kaon and pion productions in heavy-ion collisions~\cite{fuchs2001,sturm2001,hartnack2006,xiao2009}. 

With improvements in sensitivity of current gravitational-wave interferometers, future measurements of tidal deformations in binary NS mergers will provide more stringent constraints on the dense-matter EoS. 

\section*{Acknowledgments}
This work was financially supported by Fonds de la Recherche Scientifique (Belgium) under grants no. CDR J.0115.18. and no. 1.B.410.18F, and the European Cooperation in Science and Technology (COST) Action CA16214. 

\bibliography{biblio}
\end{document}